\documentclass{PoS}
\title{Resonances in unitarized HEFT at the LHC}

\ShortTitle{Resonances in unitarized HEFT at the LHC}

\author{\speaker{Antonio Dobado}, Felipe J. Llanes-Estrada and Alexandre Salas-Bern\'ardez \thanks{ Work supported by Spanish grants MINECO:FPA2014-53375-C2-1-P, FPA2016-75654-C2-1-P and the COST Action CA16108.}\\
Dept. F\'{\i}sica Te\'orica e IPARCOS, Univ. Complutense de Madrid, Plaza de las Ciencias 1, 28040 Madrid\\
E-mail: \email{dobado@fis.ucm.es}}

\author{Rafael L. Delgado \thanks{ Work supported by the INFN post-doctoral Fellowship Programme, the Ram\'on Areces Foundation and the COST Action CA16108.} \\
        INFN Sezione di Firenze, Via G. Sansone 1, I-50019 Sesto Fiorentino, Italy\\
        E-mail: \email{rdelgadol@fi.infn.it}}

\abstract{Higgs Effective Field Theory (HEFT) is deployed to study elastic vector-boson  scattering at the high LHC energies. The interaction is strong over most of the parameter space, with the minimal Standard Model being a remarkable exception. \\
One-loop HEFT complemented with dispersion relations and the Equivalence Theorem leads to two different unitarization methods which produce analytical amplitudes corresponding to different approximate solutions to the dispersion relations: the Inverse Amplitude method (IAM) and the N/D method. \\
The partial waves obtained can show poles in the second Riemann sheet whose natural interpretation is that of dynamical resonances with masses and widths depending on the starting HEFT parameters. Different unitarizations yield qualitatively, and in many cases quantitatively, very similar results. \\
The amplitudes obtained provide realistic resonant and nonresonant cross sections to be compared with and to be used for a proper interpretation of the LHC data.}

\FullConference{
European Physical Society Conference on High Energy Physics - EPS-HEP2019 -\\
			10-17 July, 2019\\
			Ghent, Belgium}

\begin{document}
The remaining key task for the LHC is to study the Electroweak Symmetry Breaking Sector (EWSBS) composed of the $W_L$ and $Z_L$ longitudinal components of the gauge bosons (which, by the Equivalence theorem, we identify at high-energy with the $\omega^i$ Goldstone bosons) and the Higgs boson $h$.
The Standard Model (SM) is the minimal theory for these four particles, and there are extensive theory studies assisting LHC efforts to extract the coefficients of its SM-EFT extension~\cite{Gomez-Ambrosio:2018pnl}.

A more general extension of the SM is the Higgs-EFT (HEFT). It does not assume that the Higgs boson belongs to a doublet representation of the electroweak gauge symmetry, and keeps the most general couplings of the new scalar particle to the Goldstone bosons of electroweak symmetry. 

It may be that, if new physics is active at a higher energy (perhaps unreachable by the LHC), the experiments are still able to measure separations of the Lagrangian parameters from their SM values. It is then a theory problem to discern what the new high-energy dynamics might be and at what energy they may appear, a key question for the planning of future endeavors.

A first option to link the low-energy HEFT to the scale of new physics is to project the underlying theory to saturate the low-energy constants by the parameters of the new resonances~\cite{Rosell:2019cto}. This procedure's poster child is the QCD $\rho$ meson. 
For example, the $l_1$ and $l_2$ parameters of the chiral Lagrangian have been extracted from precision studies of low-energy data near the pion-pion threshold~\cite{Nebreda:2012ve}
and various studies based on resonance saturation yield comparable results (see tables III and IV of~\cite{LlanesEstrada:2003ha}) if precision is not required (obviating also the difficulty of deciding at what renormalization scale should the parameters be chosen to be saturated by the resonances).
Application of this method requires at least an idea of what shape does the new physics take (since projecting from high to low energies is an information-lossy procedure).

A second option that we here pursue is to extend chiral perturbative evaluations with the low-energy parameters by employing basic theoretical principles such as unitarity and causality (encoded in dispersion relations). 
A complete description of the method has been put forward in~\cite{Delgado:2015kxa}.

The HEFT Lagrangian is an Electroweak chiral Lagrangian extended by a scalar field, and builds on the known interactions of Goldstone bosons, extending these interactions to couple the Goldstone bosons with the additional low-energy Higgs field as~\cite{Delgado:2015kxa,Delgado:2013hxa} 

\begin{eqnarray} \label{bosonLagrangian}
{\cal L} &=& \frac{1}{2}\left[1+2a\frac{h}{v}+b\left(\frac{h}{v}\right)^2\right]
\partial_\mu\omega^i\partial^\mu\omega^j\left(\delta_{ij}+\frac{\omega^i\omega^j}{v^2}\right)
+\frac{1}{2}\partial_\mu h\partial^\mu h %
\nonumber\\
 &+& \frac{4a_4}{v^4}\partial_\mu \omega^i\partial_\nu \omega^i\partial^\mu\omega^j\partial^\nu\omega^j
+\frac{4a_5}{v^4}\partial_\mu\omega^i\partial^\mu\omega^i\partial_\nu\omega^j\partial^\nu\omega^j
+\frac{g}{v^4}(\partial_\mu h\partial^\mu h)^2
\nonumber\\
 &+& \frac{2d}{v^4}\partial_\mu h\partial^\mu h\partial_\nu\omega^i\partial^\nu\omega^i
+\frac{2e}{v^4}\partial_\mu h\partial^\nu h\partial^\mu\omega^i\partial_\nu\omega^i
\ ,
\end{eqnarray}
which has seven parameters, $a$, $b$, $a_4$, $a_5$, $g$, $d$, and $e$ (see~\cite{Cata:2019edh} for a recent analysis). 

The Lagrangian of Eq.~(\ref{bosonLagrangian}) focuses on the kinematic regime for the boson-boson square CM energy $s \gg m_h^2, m_W^2$, that is, the $1\,{\rm TeV}^2$ region where derivative couplings dominate over constant couplings
such as the $c_{hhh}\frac{m_h^2}{2v} h^3$ (needed for lower-energy boson-boson production~\cite{Buchalla:2018yce}).

The generic form of the custodial isospin-angular momentum $t_{IJ}$ amplitudes derived from this Lagrangian density is an expansion in powers of $s$. They adopt the generic form
\begin{equation}
t(s) = K s + \left[B + D\log\left(\frac{s}{\mu^2}\right) + E\log\left(\frac{-s}{\mu^2}\right) \right]s^2 + \dots 
\end{equation}
where $K$, $B$, $D$ and $E$ depend on the parameters of the Lagrangian density. For example~\cite{Delgado:2015kxa,Delgado:2013hxa}, for the $\omega\omega\to hh$ process in an $s$-wave, in terms of the renormalization scale $\mu$,
\begin{eqnarray}{} \label{Mscalar}
\nonumber   K  =  \frac{\sqrt{3}}{32 \pi v^2} (a^2-b); \ \ \ \ \   
  D  =  - \frac{\sqrt{3}(a^2-b)^2}{9216\pi^3 v^4};   \ \ \  \ \ 
  E  =   -\frac{\sqrt{3}(a^2-b)(1-a^2)}{512\pi^3 v^4};  \nonumber \\
B  =  \frac{\sqrt{3}}{16\pi v^4} \left(d(\mu)+ \frac{e(\mu)}{3}\right) +\frac{\sqrt{3}}{18432\pi^3 v^4}(a^2-b)[72(1-a^2) + (a^2-b)]     
\end{eqnarray}

\begin{figure}[ht!]
\begin{minipage}{0.55\textwidth}
\centering
\includegraphics[width=0.9\columnwidth]{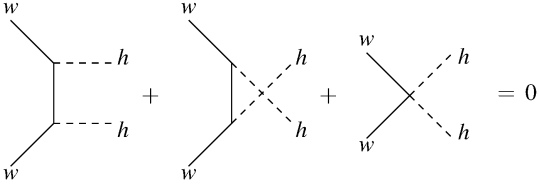}
\end{minipage}\ \ \ 
\begin{minipage}{0.38\textwidth}
\caption{The tree-level diagrams of order $s$ cancel out when the Standard Model EWSBS is cast as a HEFT, meaning there are no strong interactions among the would-be Goldstone and Higgs bosons. Here as an example the $\omega\omega \to hh$ process. \label{fig:cancel}}
\end{minipage}
\end{figure}

In the Standard Model all the NLO derivative coupling parameters vanish, particularly $d=e=0$, and the LO ones take the simple $a=1=b$ value~\footnote{This can be seen by recasting the complex doublet Higgs field of the SM in spherical coordinates $h$, $\vec{\omega}$ as $\phi^T=\left(1+\frac{h}{v}\right)\left[ \vec{\omega}, \sqrt{v^2-\vec{\omega}^2} \right]$ instead of the more usual Cartesian $\phi_1\dots\phi_4$.}; the $O(s^2)$ terms vanish directly while the $O(s)$ ones cancel because they are proportional to the $(a^2-b)$ combination, reflecting the cancellation of Figure~\ref{fig:cancel}.
Thus, the Standard Model remains renormalizable and weakly coupled even at large $s$. This is not the case for theories with new high-energy content, that reflect in deviating LO and/or NLO parameters.

The interesting question is then how should the LHC data be used to assess that new content, in case any SM deviations are found. Our answer is to construct amplitudes for the $\omega^i$ and $h$ bosons with theoretically sound properties (unitarity, order by order renormalizability and causality in the form of $s$ plane analyticity safe cuts, and agreement with the low-energy EFT when expanded at low-$s$). 
We have explored several methods (the improved-K matrix, the large-$N$ expansion, the N/D method, and the Inverse Amplitude Method) and found them to yield similar results for the high-energy scale when fed with the same low-energy parameters.

\begin{figure}[ht!]
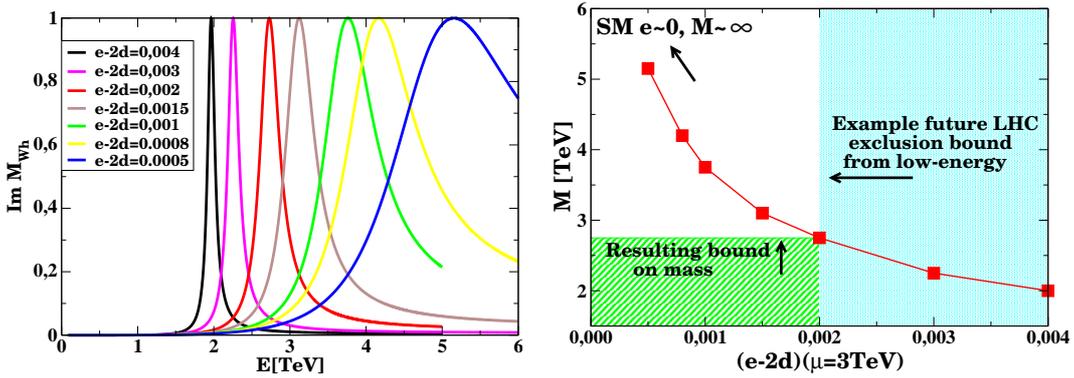

\begin{center}
\includegraphics[width=0.45\textwidth]{Resonancias1_2019.eps} \ \ \ 
\includegraphics[width=0.45\textwidth]{Resonancias2_2019.eps}
\end{center}
\caption{Left: resonances in the elastic $\omega h \to \omega h$ channel (Mandelstam-crossed from the $\omega\omega\to hh$ treated in~\cite{Delgado:2014dxa}). From left to right the mass of the new physics resonance increases, corresponding to decreasing values of the $(e-2d)$ combination of low-energy HEFT NLO parameters. Right: the resonance mass (extracted at saturation of unitarity ${\mathcal Im}\ t_{11}=1$) as function of $(e-2d)=e$ (we set $d=0$) from the left plot. As the LHC proceeds from right to left setting bounds to the parameter, which can be done from low-energy measurements at high precision, it also sets a bound on the mass-scale of the new physics. The Standard Model is recovered by sending $(e-2d)\to 0$ (with $a^2=1=b$) which removes the resonance from the spectrum, $M\to \infty$.
\label{resonances}}
\end{figure}

To exemplify, let us show one calculation for $\omega h\to \omega h$ with the IAM which, in spite of its sophisticated anchoring in dispersive theory, takes a very simple algebraic form as 
\begin{equation}
t_{\rm IAM}(s) = \frac{(t^{(0)})^2}{t^{(0)}-t^{(1)} } 
\end{equation}
built from the order $s$ and $s^2$ terms of the amplitude's perturbative expansion.
A check on the reliability of the IAM has been carried out~\cite{Corbett:2015lfa} by eliminating a heavy scalar particle from the theory and trying to reconstruct its mass (successfully) from its imprint in the low-energy parameters.
Figure~\ref{resonances} presents, on the left panel, the dependence on the $(e-2d)$ parameter combination (see~\cite{Dobado:2017lwg} for a complete analysis of this channel).

The resonances of boson-boson scattering appear as poles in the second Riemann sheet of the amplitude $t$ and, if near enough the real axis, leave a characteristic resonant shape of the scattering amplitude as visible in the figure. It is clear that the method can be used to determine the scale at which new physics enters from a measurement of the low-energy parameters. 


\end{document}